\documentclass[12pt]{article}
\usepackage{amsmath,amssymb,mathrsfs,url,graphicx,hyperref}

\title{Distribution of the Time at Which\\ an Ideal Detector Clicks}

\author{
Roderich Tumulka\footnote{Fachbereich Mathematik,
	Eberhard-Karls-Universit\"at, 
	Auf der Morgenstelle 10, 72076 T\"ubingen, Germany.
	E-mail: roderich.tumulka@uni-tuebingen.de}
}

\date{April 25, 2022}

\newcommand{\Hilbert}{\mathscr{H}}

\newcommand{\be}{\begin{equation}}
\newcommand{\ee}{\end{equation}}
\newcommand{\scp}[2]{\langle #1|#2\rangle}
\renewcommand{\Re}{\mathrm{Re}}
\renewcommand{\Im}{\mathrm{Im}}
\newcommand{\RRR}{\mathbb{R}}
\newcommand{\CCC}{\mathbb{C}}

\newcommand{\vx}{\boldsymbol{x}}

\newcommand{\vn}{\boldsymbol{n}}

\newcommand{\vX}{{\boldsymbol{X}}}
\newcommand{\vj}{{\boldsymbol{j}}}

\newcommand{\vbeta}{\boldsymbol{\beta}}
\newcommand{\Q}{\mathcal{Q}}

\newcommand{\R}{\Omega}
\newcommand{\bou}{\partial \R}

\newcommand{\prob}{\mathrm{Prob}}
\newcommand{\Z}{\mathscr{Z}}

\begin{document}
\maketitle
\begin{abstract}
We consider the problem of computing, for a detector surface waiting for a quantum particle to arrive, the probability distribution of the time and place at which the particle gets detected, from the initial wave function of the particle in the non-relativistic regime. Although the standard rules of quantum mechanics offer no operator for the time of arrival, quantum mechanics makes an unambiguous prediction for this distribution, defined by first solving the Schr\"odinger equation for the big quantum system formed by the particle of interest, the detector, a clock, and a device that records the time and place of detection, then making a quantum measurement of the record at a very late time, and finally using the distribution of the recorded time and place. This leads to the question whether there is also a practical, simple rule for computing this distribution, at least approximately (i.e., for an idealized detector). We argue here in favor of a rule based on a 1-particle Schr\"odinger equation with a certain (absorbing) boundary condition at the ideal detecting surface, first considered by Werner in 1987. We present a novel derivation of this rule and describe how it arises as a limit of a ``soft'' detector represented by an imaginary potential.

\medskip

\noindent 
Key words: detection time in quantum mechanics, time observable, time of arrival in quantum mechanics, absorbing boundary condition in quantum mechanics, quantum Zeno effect, imaginary potential, POVM.
\end{abstract}

\section{Introduction}

Consider a region
$\R \subset \RRR^3$ in physical space with detectors placed
everywhere along the boundary $\bou$, and a non-relativistic quantum particle starting
at time $t=0$ with wave function $\psi_0$ whose support lies inside
$\R$. Sooner or later, one of the detectors may register the particle,
thus defining exit time $T$ and exit position $\vX$ in $\bou$, which we
combine into the pair $Z= (T,\vX)$; in case no detector ever clicks we
write $Z= \infty$. Our goal is to predict, from $\psi_0$, the
probability distribution of the random variable $Z$ in $\Z =
[0,\infty) \times \bou \cup \{\infty\}$. Put succinctly, we are asking for a Born rule on the timelike surface $[0,\infty) \times \bou$ in non-relativistic space-time $\RRR \times \RRR^3$.

Although there is no self-adjoint time operator in the usual formalism of quantum mechanics, the distribution of $T$ and $\vX$ can in principle be computed from quantum mechanics by treating the detectors themselves as quantum mechanical systems, coupled to the particle under study, and by further including a clock and a device that records the time of the click and the location of the detector that clicked. Suppose that the full system $S$ remains isolated until a late time $t_f$; 
then the Schr\"odinger equation determines $S$'s wave function $\Psi_{t_f}$, and the distribution $|\Psi_{t_f}|^2$ determines in particular 
the probability that the record was $(t,\vx)$.
While this procedure provides no practical method of computing the distribution of $Z=(T,\vX)$, it implies that the distribution of $Z$, as a function of $\psi_0$, is of the form
\begin{equation}\label{statrealdet}
  \prob_{\psi_0} \bigl( Z \in B\bigr) = \scp{\psi_0} {E(B)| \psi_0}
\end{equation}
for some positive-operator-valued measure (POVM) $E$ on
$\Z$ (see \cite{op} or \cite[p.~6]{BGL}, and \cite{povm} about POVMs in general). 
The back effect of the presence of detectors on the
wave function is already included in \eqref{statrealdet}, and the POVM $E$ will in principle depend on 
the initial wave function of the detectors.

A practical method of computing $E$ would exist, however, if there is a POVM $E_0$, or a family $E_\kappa$ depending on one or few parameters $\kappa$, representing
an ``idealized detecting surface'' at $\bou$, such that for every setup of
real detectors along $\bou$ the true POVM $E$ is reasonably close to one of the $E_\kappa$ and the $E_\kappa$ can be defined in a simple way. (Let us call that the ideal detector hypothesis.) The idealized POVM $E_\kappa$ would disregard physical details that may vary from one detector to another, and would represent the ``best possible'' detector, which real detectors should be designed to approximate. In fact, it appears to be common experience that detection probabilities do not depend in an essential way on the physical nature of the detectors, except that different types of detectors are sensitive to different species of particles and at different energies. Moreover, the fact that the position operators of quantum mechanics define the probability distribution of the particle if we choose to detect it at time $t_0$, viz., $|\psi_{t_0}(\vx)|^2$, independently of the details of the detector, seems to support the possibility of a mathematical concept of an ideal detector.

In this paper, we argue in favor of a particular proposal of such a POVM $E_\kappa$, i.e., of a specific candidate for the ideal detector hypothesis. We propose a practical rule for computing the distribution of $Z$ that we call the \emph{absorbing boundary rule} (ABR) and that involves an \emph{absorbing boundary condition} (ABC), Eq.~\eqref{bc} below, for the Schr\"odinger equation. This boundary condition was considered before by Werner \cite{Wer87} and Fevens and Jiang \cite{FJ99}. Fevens and Jiang considered it for a different purpose;\footnote{They used it for improving algorithms for numerically solving the Schr\"odinger equation in $\RRR^d$ by removing parts of the wave function that have propagated outside the region $\R$ represented by the finitely many grid points of the algorithm; actually, they dropped the boundary condition \eqref{bc} used here in favor of a higher-order boundary condition that absorbs more of the wave.} Werner \cite{Wer87} considered it and the associated probability distribution (given by Eq.~\eqref{probnjR} below) as an example of a detection time distribution; we note, though, that the spirit of Werner's paper was different from ours: it was to regard this distribution as just a particular example while ``any contraction semigroup determines a natural arrival time observable'' \cite{Wer87}, whereas we ask here, which observable or POVM (or contraction semigroup, see below) we should use if we want to represent detectors on $\bou$. Be that as it may, the ABC \eqref{bc} did not receive much attention in the literature; for example, in an 86-pages review paper \cite{ML00} on arrival times, it was mentioned in passing but not even written down; in most works on the subject, e.g., \cite{GRT96,AOPRU98,MSE08,MRC09,VHD13,Dhar14}, it was not mentioned at all. So it may be relevant to explain why this proposal is particularly natural.\footnote{The ABC has been studied also by Dubey, Bernardin, and Dhar \cite{DBD20}, but that was after they were aware of a preprint of the present paper.}

It would be interesting to study a detailed microscopic model of a detector, but we do not do this here; rather, we explain why the ABR achieves exactly what one should look for in a candidate of a notion of ideal detector. 
The POVM $E_\kappa$ depends on, apart from the surface $\bou$ and the particle's mass $m$, a  parameter $\kappa>0$ that we call the \emph{wave number of sensitivity} of the detector. We present a derivation of the ABR from the quantum mechanics of the system $S$ formed by particle and detector, and we describe how the ABR arises as a limiting case of an ideal ``soft'' detector (i.e., one that takes a while to notice the particle) represented by an imaginary potential. 

Previous proposals for an ideal POVM $E_0(\cdot)$ took for granted that it should obey a simple relation to the free time evolution (i.e., the one that would be obtained in the absence of detectors) \cite{AB61,Kij74,Wer86,Bau00,BF02,Gal02,Gal04}; or that $E_0(\cdot)$ can be obtained by quantizing a classical expression \cite{AB61,Wer88,GRT96,Bau00,Gal02,Gal04}; or that no waves should be reflected from the detecting surface \cite{All69b,ML00}. All of these assumptions seem questionable, and even more so in the light of the derivation we present here. Moreover, previous proposals were not backed up by an analysis of the quantum mechanical workings of a detector, except for soft detectors \cite{All69b,HSM03,RMH09}.

Two basic properties of the ABR are that 
\begin{itemize}
\item[(i)] it describes a ``hard'' detector, i.e., one that detects a particle \emph{immediately} when it arrives at the surface $\bou$, and
\item[(ii)] a wave packet moving towards the detecting surface will not be completely absorbed but partly reflected. 
\end{itemize}
We will explain why one should expect property~(ii), and why (i) and (ii) do not exclude each other. (In a sense, the boundary absorbs the particle but not completely the wave.) In view of the quantum Zeno effect \cite{Fri72,Zeno,Dhar13,Dhar14}, it may seem surprising that a rule can have property (i). 
Elsewhere, we will describe the natural extension of the ABR to the case of moving detectors, that of several particles, that of particles with spin, and that of the Dirac equation, as well as derive an uncertainty relation between the detection time $T$ and the energy of the initial wave function $\psi_0$. 
Dubey, Bernardin, and Dhar \cite{DBD20} have shown that the ABR can be obtained in a limit similar but not identical to that considered in the quantum Zeno effect, involving repeated quantum measurements of the projection to $\R$ at time intervals of length $\tau$, on a particle moving on a lattice of width $\varepsilon$; this limit involves first letting $\tau\to0$ while assuming the transition amplitudes in the Hamiltonian between the lattice sites of $\bou$ and their neighbors in the interior of $\R$ diverge like $\tau^{-1/2}$, and then letting $\varepsilon\to 0$ while rescaling time and $\psi$ appropriately.

\section{Statement of the Absorbing Boundary Rule}

Let $\kappa>0$ be a constant of dimension 1/length. Solve the  Schr\"odinger equation
\be\label{Schr}
i\hbar\frac{\partial \psi}{\partial t} = -\frac{\hbar^2}{2m} \nabla^2 \psi + V\psi
\ee
in $\R$ with potential $V:\R\to\RRR$ and boundary condition
\be\label{bc}
\frac{\partial \psi}{\partial n}(\vx) = i\kappa \psi(\vx)
\ee
at every $\vx\in\bou$, with $\partial/\partial n$ the outward normal derivative on the surface, 
$\partial \psi/\partial n := \vn(\vx) \cdot \nabla\psi(\vx)$
with $\vn(\vx)$ the outward unit normal vector to $\bou$ at $\vx\in\bou$. (As we will describe elsewhere, it follows from the Hille--Yosida theorem that a solution to \eqref{Schr} and \eqref{bc} exists and is unique for $\psi_0\in L^2(\R)$, a result also obtained in \cite{Wer87}.)
Assume that $\|\psi_0\|^2 = \int_\R d^3\vx\, |\psi_0(\vx)|^2=1$.
Then, the rule asserts,
\begin{equation}\label{probnjR}
  \prob_{\psi_0} \Bigl( t_1 \leq T<t_2, \vX \in B \Bigr) =
  \int\limits_{t_1}^{t_2} dt \int\limits_{B} d^2\vx \; \vn(\vx) \cdot
  \vj^{\psi_t}(\vx)
\end{equation}
for any $0\leq t_1<t_2$ and any set $B\subseteq \bou$, with $d^2\vx$ the surface area element and $\vj^\psi$ the probability current vector field
defined by $\psi$, which is
\begin{equation}\label{jSchr}
  \vj^\psi = \frac{\hbar}{m} \Im\, \psi^* \nabla \psi\,.
\end{equation}
In other words, the joint probability density of $T$ and $\vX$ relative to $dt\, d^2\vx$ is the normal component of the current across the boundary, $j_n^{\psi_t}(\vx)=\vn(\vx)\cdot \vj^{\psi_t}(\vx)$. Furthermore,
\be\label{Zinfty}
\prob_{\psi_0}(Z=\infty) = 1-  \int\limits_{0}^{\infty} dt \int\limits_{\bou} d^2\vx \; \vn(\vx) \cdot
  \vj^{\psi_t}(\vx)\,.
\ee
This completes the statement of the rule.

\section{Properties of the Absorbing Boundary Rule}

\noindent{\bf Proposition 1.} {\it The ABR defines a probability distribution.}

\bigskip

\noindent{\it Derivation.} We note that $\vj^\psi$ is always outward-pointing at the boundary as a consequence of the boundary condition \eqref{bc}:
\be
j_n^\psi(\vx) 
=  \frac{\hbar}{m} \Im \Bigl[\psi^*(\vx)\, \frac{\partial \psi}{\partial n}(\vx)\Bigr]
=  \frac{\hbar\kappa}{m} |\psi(\vx)|^2 \geq 0\,.
\ee
This density, when integrated over $[0,\infty)\times\bou$, cannot be greater than 1 because $\|\psi_0\|^2=1$ and
\be\label{eq1}
\int_0^\infty dt \int_{\bou} d^2\vx \, j_n^\psi(\vx,t)
= \int_{\R} d^3\vx\, |\psi_0(\vx)|^2 - \lim_{t\to\infty} \int_{\R} d^3\vx\, |\psi_t(\vx)|^2 \,,
\ee
an equation that follows from the continuity equation implied by the Schr\"odinger equation, $\partial |\psi|^2/\partial t = -\nabla\cdot \vj^\psi$, by integrating over $\R$, applying the divergence theorem, and integrating over $t$. In particular, the right-hand side of \eqref{Zinfty} is non-negative and, in fact, equal to $\lim_{t\to\infty} \|\psi_t\|^2$. It follows also that $\|\psi_t\|^2=\int_{\R}d^3\vx\, |\psi_t(\vx)|^2$ is not conserved but instead is a decreasing function of $t$, so the time evolution of $\psi_t$ in $\Hilbert=L^2(\R)$ is not unitary. In fact, $\|\psi_t\|^2$, rather than being
1, is the probability that $T>t$ or $Z=\infty$, i.e., that no detection has occurred up to time $t$.\hfill$\square$

\bigskip

The rule corresponds to a POVM $E_\kappa$ that can be expressed as
\begin{align}
  E_\kappa \bigl( dt\times d^2\vx \bigr) 
  &= \frac{\hbar\kappa}{m} \, W_t^\dagger \,
  |\vx\rangle\langle\vx|  \,W_t\,  dt\, d^2\vx\\
E_\kappa (\{\infty\}) 
&= I-E_\kappa([0,\infty)\times \bou)
=\lim_{t\to\infty} W_t^\dagger W_t 
\end{align}
with ${}^\dagger$ denoting the adjoint operator, $I$ the identity operator, and $W_t$ the (non-unitary) linear operator that maps $\psi_0$ to $\psi_t$ solving \eqref{Schr} and \eqref{bc} for $t\geq 0$. The $W_t$ have the properties $W_0=I$, $W_t W_s=W_{t+s}$, and $\|W_t\psi\|\leq \|\psi\|$; that is, they form a \emph{contraction semigroup}. Since the $E_\kappa(dt)$ are not projections, there are no eigenstates of detection time.

These considerations can be visualized in terms of Bohmian trajectories $\vX(t)$ \cite{DT}: Bohm's equation of motion,
\be
\frac{d\vX(t)}{dt} = \frac{\vj^{\psi_t}(\vX(t))}{|\psi_t(\vX(t))|^2}\,,
\ee
implies together with \eqref{bc} that trajectories can cross the boundary $\bou$ only outwards, in fact with the prescribed normal velocity $\hbar\kappa/m$. A detector clicks when and where the Bohmian particle reaches $\bou$; the probability distribution of this space-time point agrees with \eqref{probnjR} since the initial distribution of the Bohmian particle is $|\psi_0|^2$ \cite{Daumer}. 
If we had solved the Schr\"odinger equation on $\RRR^3$ instead of $\R$, and without the boundary condition \eqref{bc}, as would be appropriate in the absence of detectors, then the Bohmian trajectory might cross $\bou$ several times, re-entering $\R$ after having left it \cite{gruebl,KGE03,VHD13}. The boundary condition \eqref{bc} excludes this. This example also illustrates why ``time of detection'' is a more accurate name for $T$ than ``time of arrival.''

\begin{figure}[h]
\begin{center}
\includegraphics[width=.6 \textwidth]{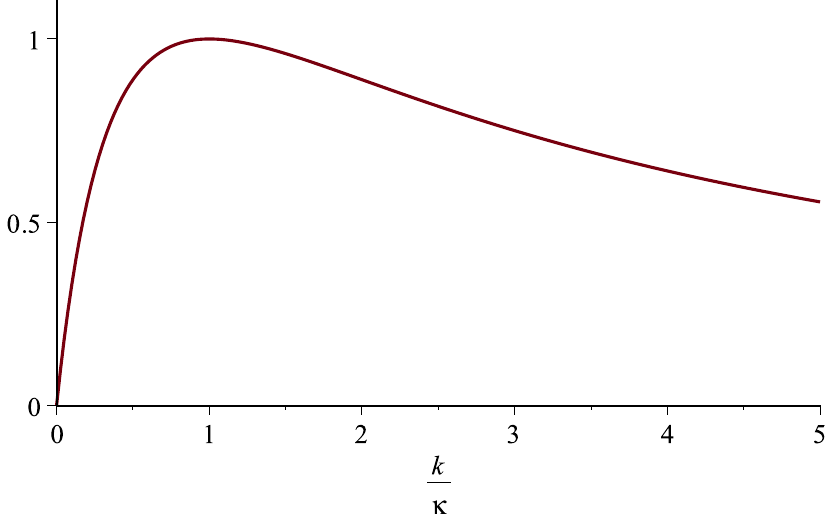}
\end{center}
\caption{Graph of the absorption strength $A_k$ of the ideal detecting surface as a function of wave number $k$ in units of $\kappa$. The maximum attained at $k=\kappa$ is equal to 1, corresponding to complete absorption.}
\label{fig:1-c}
\end{figure}

Although all Bohmian trajectories that reach $\bou$ have to cross it, part of the wave function reaching $\bou$ can be reflected. 
To quantify the reflection, consider the 1-dimensional version of the absorbing boundary rule for $\R=(-\infty,0]$ with boundary $\bou=\{0\}$ and boundary condition $\psi'(0)=i\kappa\psi(0)$. 
The reflection coefficient $R_k$ at wave number $k$ is given by $R_k=|c_k|^2$, where $c_k$ is the complex coefficient ensuring that the eigenfunction $\psi(x) = \exp(ikx) + c_k\exp(-ikx)$ 
satisfies the boundary condition; one finds that  
\be\label{R}
R_k = (k-\kappa)^2/(k+\kappa)^2\,.
\ee
The absorption coefficient is $A_k=1-R_k$, whose graph is depicted in Figure~\ref{fig:1-c}. At $k=\kappa$, the wave is completely absorbed, while waves of other wave numbers are partly absorbed and partly reflected. This means that our ideal detector surface absorbs (and detects) well in a certain energy range but poorly at much higher or lower energies (in agreement with the results of \cite{AOPRU98}), and $\kappa$ is the wave number at which the detector is most sensitive.

\section{Derivation of the Absorbing Boundary Rule}

Regard the detectors as a quantum system $D$ with configuration space $\Q_D=\RRR^{3N}$; call the particle $P$, with configuration space $\Q_P=\RRR^3$; the whole system $S=P\cup D$, with configuration space $\Q_S=\Q_P\times \Q_D$, evolves unitarily with initial wave function $\Psi_0=\psi_0\otimes \phi_0$. Let $\Xi_D$ denote the set of $D$-configurations in which the detectors have not clicked but are ready, and $\Upsilon_D$ the set of those in which a detector has fired, so the initial wave function of $D$, $\phi_0$, is concentrated in $\Xi_D$, and $\Psi_0$ is concentrated in $\R\times \Xi_D\subset \Q_S$. The interaction between $P$ and $D$ occurs, not in the interior of $\R\times \Xi_D$, but only at the boundary $\bou\times\Xi_D$: Any probability current in $\Q_S$ that reaches $\bou\times\Xi_D$ should be transported quickly (ideally, immediately) to $\bou\times\Upsilon_D\subset \Upsilon_S:=\Q_P\times \Upsilon_D$, a region of $S$-configuration space that is far from $\bou\times\Xi_D$, as configurations in $\Upsilon_S$ are macroscopically different from those in $\Xi_S=\Q_P\times \Xi_D$. Due to this separation, parts of $\Psi$ that have reached $\Upsilon_S$ should not be able to propagate back to $\Xi_S$ and interfere there with parts of $\Psi$ that have not yet left $\Xi_S$; that is, the detection is practically irreversible, resulting in \emph{decoherence} between the part of the wave function that has passed $\bou$ and the part that has not. 
Also, the probability current should always flow from the interior of $\R\times \Xi_D$ to $(\Q_P\setminus \R)\times \Upsilon_D$. As a consequence, the $P$-component of the current at $\bou\times\Xi_D$ should point outward. We are thus led to the following picture: 
\begin{itemize}
\item[(i)] The Schr\"odinger equation \eqref{Schr} holds for $\psi$ inside $\R$. 
\item[(ii)] Something happens on $\bou$, which should not depend sensitively on the details of the initial detector state $\phi_0$. 
\item[(iii)] The evolution of $\psi_t$ in $\R$ is still linear, but no longer unitary because $\psi_t$ corresponds to only a part of the full wave function $\Psi_t$, i.e., the part in $\Xi_S$. 
\item[(iv)] The current $\vj^{\psi_t}(\vx)$ at $\vx\in\bou$ always points outward, $\vn(\vx)\cdot \vj^{\psi_t}(\vx)\geq 0$. 
\item[(v)] The evolution of $\Psi_t$ in $\Xi_S$ is autonomous, i.e., not affected by whatever $\Psi_t$ looks like in $\Upsilon_S$, as those parts cannot propagate back to $\Xi_S$. 
\item[(vi)] Thus, the evolution of $\psi_t$ in $\R$ should be autonomous, depending only on few parameters (``$\kappa$'') encoding properties of the detectors. 
\end{itemize}

These features suggest a boundary condition at $\bou$ for $\psi_t$. By (iii), the boundary condition should be linear, and since the Schr\"odinger equation involves second-order space derivatives, it may involve derivatives up to first order. The boundary condition should be local, i.e., involve only one $\vx$, as there is nothing in the setup that would connect several boundary points. The most general boundary condition of this kind is
\be
\alpha(\vx) \psi(\vx) + \vbeta(\vx) \cdot \nabla\psi(\vx) =0
\ee
for $\vx\in\bou$ with coefficients $\alpha(\vx)\in\CCC,\vbeta(\vx)\in\CCC^3$, known as a \emph{Robin boundary condition}. It entails the following condition on the current $\vj$ (multiply by $(\hbar/m)\psi^*(\vx)$ and take the imaginary part):
\be
(\Re\, \vbeta) \cdot \vj = -\tfrac{\hbar}{m} (\Im\, \alpha) |\psi|^2 + \tfrac{\hbar}{m} (\Im\, \vbeta)\cdot\Re(\psi^* \nabla \psi)\,.
\ee
This condition will enforce that the current points outward, $j_n\geq 0$, if and only if $\Re\, \vbeta(\vx)=\gamma(\vx)\,\vn(\vx)$ with $0\neq\gamma(\vx)\in\RRR$, $\Im\, \vbeta=0$, and $\gamma^{-1}\Im\, \alpha \leq 0$. In case $\Im\, \alpha=0$, it forces the normal current to vanish, so the relevant conditions (for which sometimes $j_n>0$) are those with $\gamma^{-1}\Im\, \alpha<0$. 

We are thus led to a generalized version of \eqref{bc} in which $i\kappa$ is replaced by $\nu+i\kappa$ with $\kappa>0$ and $\nu\in \RRR$. It is then still the case that \eqref{probnjR} defines a probability distribution on $\Z$, but the maximal absorption (which still occurs at $k=\kappa$) is strictly less than 1 if $\nu\neq 0$, which suggests that the detector represented by this boundary condition is in a sense less than perfect. Also, a non-zero $\nu$ leads to a complex coefficient $c_k$ for the reflected wave in $\psi(x) = \exp(ikx) + c_k\exp(-ikx)$, so that the reflected wave undergoes a phase shift, which may occur for a real detector but complicates matters unnecessarily if we want to consider an \emph{ideal} detector. As mentioned, desideratum (vi) is satisfied: the time evolution of $\psi_t$ is autonomous (also if $\nu\neq 0$).

In the Bohmian picture it becomes particularly transparent why one should demand that the current points outward on $\bou$ but not that the reflection coefficient vanishes. That is because the Bohmian picture provides a clear distinction between the absorption of the \emph{particle} at $\bou$ and the absorption of the \emph{wave}. 
When the particle reaches $\bou$ then detection should be inevitable and irreversible. 
There is no reason, however, to expect that the wave arriving at $\bou$ should be absorbed completely. On the contrary, one should expect partial reflection, so that the presence of the detector influences the evolution inside $\R$.

\section{Soft Detectors}

The absorbing boundary rule describes ``hard'' detectors. Real detectors, in contrast, may be ``soft.'' A mathematical description of a soft detector, on a par with the absorbing boundary rule, can be given in terms of imaginary potentials, i.e., by adding $-iv$ to the Hamiltonian at every $\vx$ in the detector volume $\R_D$, where $v>0$ is a constant, the detection rate. Complex potentials have been used by various authors to model the absorption or removal of a quantum particle \cite{Bet40,MM65,Lev69,All69b,Hod71}. The probability distribution of the detection time and place, $Z=(T,\vX)\in [0,\infty)\times \R_D$ or $Z=\infty$, is given by
\be
\prob_{\psi_0}\Bigl( t_1 \leq T<t_2, \vX \in B \Bigr) =
  \int\limits_{t_1}^{t_2} dt \int\limits_{B} d^3\vx \; v|\psi_t(\vx)|^2
\ee
for any $B\subseteq \R_D$.
Again, $\|\psi_t\|^2$ equals the probability that $T>t$ or $Z=\infty$. The Bohmian particle, whenever in $\R_D$, gets absorbed at rate $v$. 

Suppose that the detector volume $\R_D$ is a shell of thickness $L>0$ around $\R$, and let $\R_L=\R\cup \R_D$ be the neighborhood of radius $L$ around $\R$. At the outermost surface $\bou_L$, we impose a Neumann boundary condition, $\partial\psi/\partial n=0$. As we will discuss elsewhere in detail:

\bigskip

\noindent{\bf Proposition 2.} {\it In the limit $v\to \infty$, $L\to 0$, $0<\lim(vL)<\infty$, the time evolution of $\psi$ and the distribution of $Z$ approach those of the absorbing boundary rule with
\be
\kappa = \frac{2m}{\hbar^2} \lim(vL)\,.
\ee
This is still true if the Neumann condition is replaced by a Robin boundary condition $\partial \psi/\partial n = c\, \psi$ with arbitrary real constant $c$, but not if replaced by a Dirichlet condition, $\psi=0$ on $\bou_L$.}

\bigskip

While previous authors such as Allcock \cite{All69b} have also considered representing a soft detector by an imaginary potential, they have not managed to identify a limit in which a non-trivial theory of a hard detector arises. In fact, Allcock considered the limit $v\to\infty$, $L=\infty$ (without any boundary condition) and obtained correctly that the limit is equivalent to the Schr\"odinger equation with a Dirichlet boundary condition at $\bou$, which has zero current into the boundary and thus zero probability that the detector would ever be triggered. (Allcock concluded incorrectly that a concept of an ideal hard detector is impossible.)

In scattering theory, the scattering cross-section represents the probability distribution of where the particle gets detected on a surface $\bou$ that is a sphere of radius $r$ in the limit $r\to\infty$. One assumes that no part of the arriving wave gets reflected, $R_k=0$ for all $k$. As we will discuss elsewhere, for the ideal soft detectors described above, $R_k\to 0$ for all $k$ in the limit in which $v\to0$ and $vL\to \infty$---a limit of infinite softness that seems entirely admissible in the scattering regime because if we allow very large distances $r\to\infty$ from the scattering center and very large times $t$ then we may as well allow large distances and times within the detector volume before the particle gets detected.

\section{Conclusions}

We have argued that the concept of an ideal detecting surface $\bou$ is naturally implemented in quantum mechanics by means of a certain absorbing boundary condition (ABC) and have formulated the corresponding rule for computing the joint probability distribution of detection time $T$ and detection place $\vX\in\bou$ (absorbing boundary rule, ABR). The ABR can be regarded as an analog of Born's rule for timelike surfaces in space-time. We have explained why one should expect the behavior of detectors to lead to this rule and have discussed properties of the ABC such as: that the probability distribution is well defined and given by a POVM; that the effective 1-particle Schr\"odinger evolution with the ABC is non-unitary and involves a Hamiltonian that is not self-adjoint; why the wave can be partly reflected although the particle is sure to be absorbed; that the back effect of the presence of the detectors on the particle is already taken into account; and how the ABR is related to the representation of soft detectors by means of imaginary potentials.

\medskip

\noindent \textit{Acknowledgments.} I am indebted to Abhishek Dhar, Detlef D\"urr, Shelly Goldstein, Helmut Br\"uchle, and Stefan Teufel for helpful discussions.

\end{document}